\documentclass[twocolumn,seceq]{jpsj3}
\usepackage{graphicx}

\title{Hyperbolic Deformation Applied to $S = 1$ Spin Chains\\
--- Scaling Relation in Excitation Energy  ---}

\author
{Hiroshi {\sc Ueda}$^1$, Hiroki {\sc Nakano}$^2$,
Koichi {\sc Kusakabe}$^1$, and Tomotoshi {\sc Nishino}$^3$ }

\inst
{
$^{1)}$ Department of Material Engineering Science, 
Graduate School of Engineering Science, 
Osaka University, Toyonaka 560-8531\\
$^{2)}$ Graduate School of Material Science, 
University of Hyogo,
Kamigori-cho, Hyogo 678-1297\\
$^{3)}$ Department of Physics, 
Graduate School of Science, 
Kobe University, Kobe 657-8501
}

\recdate{\today}

\abst
{
We investigate excitation energies of hyperbolically deformed $S = 1$ 
spin chains, which are specified by the local energy scale $f_j^{~} = \cosh j \lambda$, 
where $j$ is the lattice index and $\lambda$ is the deformation parameter. 
The elementary excitation is well described by a quasiparticle hopping model, which is 
also expressed in the form of hyperbolic deformation. 
It is possible to estimate the excitation gap $\Delta$ 
in the uniform limit $\lambda \rightarrow 0$, by means of a finite size scaling 
with respect to the system size $N$ and the deformation parameter $\lambda$.
}
\kword{DMRG, scaling analysis, excitation gap, spin-wave velocity, Heisenberg chain, Haldane gap}

\begin{document}
\sloppy
\maketitle

\section{Introduction}

A role of numerical study in condensed matter physics
is to analyze ground state properties and also low-energy excitations in the 
thermodynamic limit. Precise determination of the excitation gap is important, 
since presence of a finite excitation gap corresponds to the finitely correlated
property of the system. In numerical analyses of excitation gap, 
the finite size scaling~\cite{FSS1, FSS2} has been employed. This is because
computational resources are limited, and therefore the direct treatment of
infinite system is occasionally difficult.~\cite{iTEBD}

Density matrix renormalization group (DMRG) method is widely applied to one-dimensional
 (1D) quantum systems,~\cite{White1,White2,Peschel,Schollwoeck} since the method
provides precise low-lying eigenvalues of large scale systems, up to hundreds or
even thousands of sites. Although it is possible to apply DMRG method to systems with
periodic boundary conditions (PBC),~\cite{White2,Everts,Verstraete} a majority 
of applications are performed under the open boundary conditions (OBC). This is 
because numerical implementation is much easier with the use of OBC. However, 
presence of boundary corrections often makes precise 
scaling analyses difficult. To suppress such a corrections, 
so called the smooth boundary condition has been applied.~\cite{vekic:PRL71,
vekic:PRB53,Andrej:PTP122}. It is reported that an adiabatic decay of interaction 
strength toward the system boundary drastically decreases the boundary effect
on one-point functions, such as the energy density.

The smooth boundary condition is not applicable when one is interested in elementary 
excitations. This is because presence of a small energy scale near the
boundary induces a fictitious excitation localized around the boundary. 
We therefore considered to the opposite direction, and proposed the hyperbolic 
deformation;~\cite{Ueda,Ueda:PPT124} we introduced position dependence
to 1D lattice Hamiltonians, where the energy scale {\it increases} toward the both ends of the 
system. As a consequence of this non-uniform deformation, the
excited quasiparticle is weakly bounded around the center of the system. The boundary 
effect for the excitation energy is reduced, since the quasiparticle does not reach the
boundary. 

In the previous study, we chiefly applied the hyperbolic deformation to
the free Fermionic lattice model, and showed that a two-parameter scaling function exists with 
respect to the system size $N$ and the deformation parameter $\lambda$.~\cite{Ueda:PPT124} 
It is expected that a wide class of 1D systems under the hyperbolic deformation obeys 
this kind of two-parameter FSS, which could precisely estimate the excitation gap.
In this article we investigate the efficiency of this FSS in the context of hyperbolic deformation, when
it is applied to correlated systems, such as the $S=1$ AKLT chain~\cite{Affleck:PRL59} 
and the $S=1$ Heisenberg chain with uniaxial anisotropies (XXZ+D). In the next section we 
introduce hyperbolically deformed Hamiltonians for these systems. In \S 3 the form of the two-
parameter FSS is shortly reviewed. Numerical results on the spin chains are shown in \S 4. 
Conclusions are summarized in the last section.

\section{Hyperbolic Deformation}

Let us consider a group of 1D quantum spin chains, 
which are characterized by the non-uniform Hamiltonian
\begin{equation}
\hat{H}^{}_{}(\lambda) =
\sum_{j=-L/2+1}^{L/2-1} \cosh j \lambda  \,\,\, \hat{h}^{}_{j,j+1} \, ,
\end{equation}
where $j$ represents the lattice index, $L$ the system size, and $\lambda$ a 
nonnegative {\it deformation} parameter. The operator $\hat{h}^{}_{j,j+1}$ 
specifies the interaction between neighboring sites. An example is the 
bilinear-biquadratic interaction
\begin{equation}
\hat{h}^{}_{j, j+1} = 
J \cos\theta \,\,\, \hat{{\bf S}}_{j} \cdot \hat{{\bf S}}_{j+1} + 
J \sin\theta \left( \hat{{\bf S}}_{j} \cdot \hat{{\bf S}}_{j+1} \right)^2_{} 
\end{equation}
between neighboring $S = 1$ spins, where $J > 0$ is the coupling constant, 
and where the angle $\theta$ determines the ratio between bilinear and
biquadratic interactions. When $\lambda > 0$, the relative bond strength
$f_j^{~} = \cosh j \lambda$ is position dependent, and 
the weakest bond is located at the center of the system $j = 0$. 
The Hamiltonian ${\hat H}( \lambda )$ when $\lambda > 0$ can be regarded as 
a deformation --- the {\it hyperbolic deformation} --- from the uniform 
one ${\hat H}( \lambda = 0 )$.

The case $\tan \theta = 1/3$ and $J = 1 / \cos\theta$ in Eq.~(2.2) corresponds
to the AKLT interaction.~\cite{Affleck:PRL59, Affleck:CMP115}
In this case, the deformed Hamiltonian is explicitly written as
\begin{eqnarray}
&& \!\!\!\!\!\!\!\!\!\!\!\!\!\!\!\!  {\hat H}_{~}^{\rm AKLT}( \lambda ) = \\
&& \!\!\!\!\!\!\!\! \sum_{j=-L/2+1}^{L/2-1} \cosh j \lambda  
\left[
\hat{{\bf S}}_{j} \cdot \hat{{\bf S}}_{j+1} + 
\frac{1}{3} \left( \hat{{\bf S}}_{j} \cdot \hat{{\bf S}}_{j+1} \right)^2_{} 
\right] \, ,
\nonumber
\end{eqnarray}
which can be called as the deformed AKLT chain. 
Despite of the position dependence in ${\hat H}_{~}^{\rm AKLT}( \lambda )$, the 
mechanism of the $\mathbb{Z}_2 \times \mathbb{Z}_2$ symmetry 
breaking~\cite{Kennedy:PRB45} is preserved, since the Hamiltonian can be
written as a linear combination of local projections. The 
corresponding ground state is the uniform valence-bond-solid state.

A similar uniformity in the ground state is observed for the hyperbolically 
deformed $S = 1$ antiferromagnetic Heisenberg spin chain, whose Hamiltonian 
\begin{equation}
{\hat H}_{~}^{\rm AFH}( \lambda ) =  \sum_{j=-L/2+1}^{L/2-1} \cosh j \lambda  \,\, 
{\hat {\bf S}}_j^{~} \cdot {\hat {\bf S}}_{j+1}^{~}
\end{equation}
corresponds to the choice with $J = 1$ and $\theta = 0$ in Eq.~(2.2). Even in the case $\lambda > 0$ 
the expectation value $\langle {\hat {\bf S}}_j^{~} \cdot {\hat {\bf S}}_{j+1}^{~} \rangle$ 
calculated for the ground state is nearly position independent.~\cite{Ueda:arxiv0812} 
Also the deformed transverse-field-Ising model, which is defined by the Hamiltonian
\begin{equation}
{\hat H}_{~}^{\rm TFI}( \lambda ) = \sum_{j=-L/2+1}^{L/2-1} 
\cosh j \lambda  
\left[ \Gamma s_j^x + 4 s_j^z s_{j+1}^z + \Gamma s_{j+1}^z  \right] \, ,
\end{equation}
where ${\bf s}_j^{~} = ( s_j^x, s_j^y, s_j^z )$ represents the $S = 1/2$ spin at the $j$th site, 
exhibits a similar uniformity deep inside the system, 
even at the critical point $\Gamma = 1$.~\cite{Ueda:arxiv1008} These uniform 
property can be qualitatively explained by the path-integral representation 
of imaginary time evolution ${\hat U} = \exp\left[ - \tau {\hat H}( \lambda ) \right]$
under the hyperbolic geometry.~\cite{Ueda:PPT124}

\section{Scaling Form for the Excitation Energy} 

Let us consider the infinitely long and uniform $S = 1$ Heisenberg spin chain, which corresponds to 
the double limit $\lambda \rightarrow 0$ and $L \rightarrow \infty$ 
of ${\hat H}^{\rm AFH}_{~}( \lambda )$ in Eq.~(2.4), 
as a reference system. The model has finite excitation energy $\Delta$ from the 
singlet ground state, which is known as the Haldane gap.~\cite{Haldane, Haldane2}
It was reported that the quasiparticle picture well holds for the magnetic excitation.~\cite{Huse} 
When only a quasiparticle is present, the energy dispersion of the quasiparticle can be 
described by the effective Hamiltonian
\begin{equation}
{\hat H}^{\rm eff}_{~} =
- t \sum_j^{~} 
\left( 
{\hat c}^{\dagger}_{j} {\hat c}^{~}_{j+1} + 
{\hat c}^{\dagger}_{j+1} {\hat c}^{~}_{j} 
\right) 
+ ( 2 t + \Delta ) \sum_j^{~} {\hat c}^{\dagger}_{j} {\hat c}^{~}_{j} 
\end{equation}
in the vicinity of zero momentum, 
where $t$ is the hopping parameter, and where ${\hat c}^{\dagger}_{j}$ and 
${\hat c}^{~}_{j}$, respectively, represent creation and annihilation of the 
quasiparticle. The eigenvalue of the zero-quasiparticle vacuum is trivially zero, 
and that of the one-quasiparticle state is larger than zero when $\Delta$ is positive.
If more than two quasiparticles are present, interaction terms should be included 
into the effective Hamiltonian.~\cite{Schollwock:PRB53, Okunishi:PRB59}

We conjecture that such a quasiparticle picture also holds for 
the hyperbolically deformed Hamiltonian ${\hat H}^{\rm AFH}_{~}( \lambda )$ in Eq.~(2.4), 
and that the effective Hamiltonian can be written as
\begin{eqnarray}
&& \!\!\!\!\!\!\!\!\!\!\!
{\hat H}^{\rm eff}_{~}( \lambda ) = \sum_{j=-L/2+1}^{L/2-1} \cosh j \lambda 
\biggl[ - t
\left( 
{\hat c}^{\dagger}_{j} {\hat c}^{~}_{j+1} + 
{\hat c}^{\dagger}_{j+1} {\hat c}^{~}_{j} 
\right) \nonumber\\
&&~~~~~~~~ + ( 2t + \Delta ) 
\left( 
{\hat c}^{\dagger}_{j} {\hat c}^{~}_{j} + 
{\hat c}^{\dagger}_{j+1} {\hat c}^{~}_{j+1} 
\right) \biggr] 
\end{eqnarray}
in the form of hyperbolic deformation.

As a preparation for checking the correspondence between ${\hat H}^{\rm AFH}_{~}( \lambda )$ 
and ${\hat H}^{\rm eff}_{~}( \lambda)$,
the authors investigated one-particle state of ${\hat H}^{\rm eff}_{~}( \lambda )$
in our previous study.~\cite{Ueda:PPT124} The lowest-energy one-particle 
state is a shallow bound state, where the quasiparticle does not reaches to
the both end of the system when $\lambda$ is sufficiently large.
The lowest eigenvalue, which we denote as $\Delta_L^{~}( \lambda )$, 
converges to $\Delta$ in the limit $L \rightarrow \infty$ and $\lambda \rightarrow 0$.
We found that the correction $\Delta_L^{~}( \lambda ) - \Delta$ 
satisfies the two-parameter scaling 
\begin{equation}
(L+1)^2 \,\,
\frac{\Delta_{L}^{~}(\lambda) - \Delta}{t} = 
g\left[ \sqrt{ \frac{\Delta}{t}} \, (L+1)^2_{~} \, \lambda \right]
\end{equation}
with respect to the system size $L$ and the deformation parameter $\lambda$. The scaling 
function $g[ y ]$ satisfies $g[0] = \pi^2$ and $g[y \gg 1] \sim y/\sqrt{2}$. 
We conjecture that this scaling form is also applicable for elementary excitation of 
gapped spin chains in general.

\section{Two-parameter Scaling Analysis}

We perform numerical analysis of the hyperbolically deformed spin chains defined
in \S 2, by means of the DMRG method. In order to avoid the quasi four-fold 
degeneracy,~\cite{Kennedy:PRB45} we put $S = 1/2$ spins at the both ends of the 
system;~\cite{Huse} these boundary spins are not counted when we refer to the system size 
$L$. For the bond between boundary $S = 1/2$ spin and the neighboring $S = 1$ spin, 
we set weaker interaction strength $J_{\rm end}^{~} = 0.5088$. Actually the value of 
$J_{\rm end}^{~}$ is not relevant to the elementary excitation and its energy if $\lambda L$ is
sufficiently large. We keep at most $m = 100$ block spin states in the DMRG 
calculations. Numerical convergence in finite system sweeping is accelerated by use 
of the wave function prediction 
method.~\cite{White:PRL77,McClloch:arxiv0804,Ueda:JPSJ77,Ueda:JPSJ79}
We explain numerical details in the appendix A. 

\begin{figure}
\begin{center}
\includegraphics[width=80mm]{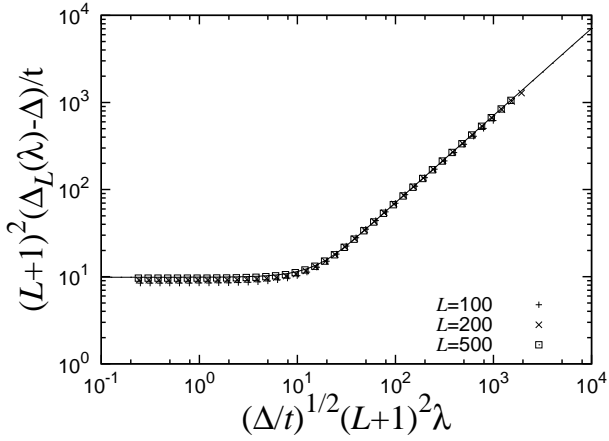}
\end{center}
\caption{Scaling plot for the hyperbolically deformed $S = 1$ 
Heisenberg chain. Best fit is realized when $\Delta=0.410485$ and $t=7.381$.
}
\end{figure}

Concerning to the ground state of the hyperbolically deformed $S = 1$ spin chains that 
we have introduced, the $z$-component of the total spin is zero. 
Let us denote the corresponding ground state energy by $E_L^{(0)}( \lambda )$. 
We also calculate the lowest eigenvalue $E_L^{(1)}( \lambda )$ when the $z$-component 
of the total spin is one.  The magnetic excitation energy is then expressed as their difference 
\begin{equation}
\Delta_L^{~}( \lambda ) = E_L^{(1)}( \lambda ) - E_L^{(0)}( \lambda ) \, .
\end{equation}
We have used the same notation $\Delta_L^{~}( \lambda )$ that appears in Eq.~(3.3), 
since we expect that $\Delta_L^{~}( \lambda )$ in Eq.~(4.1) also satisfies the scaling 
relation in Eq.~(3.3). 

For $H^{\rm HAF}_{~}( \lambda )$ in Eqs.~(2.4), we confirm 
the presence of scaling function $g$ under the choice $\Delta=0.410485$ 
and $t=7.381$ after some trials of determining these parameters.
Figure 1 shows the scaling result for 
$\Delta_{L}^{}(\lambda)-\Delta$ when $L = 100$, $200$, and $500$. These data
agrees with $\Delta_{L}^{}(\lambda)-\Delta$ shown by solid curve, which is drawn
from the effective one-particle model ${\hat H}^{\rm eff}_{~}( \lambda )$ in Eq.~(3.2).
We also calculate $\Delta_{L}^{}(\lambda)-\Delta$ for the deformed AKLT chain 
defined by ${\hat H}^{\rm AKLT}_{~}( \lambda )$ 
in Eq.~(2.3), and show the result in Fig.~2. For this case the best fit to 
${\hat H}^{\rm eff}_{~}( \lambda )$ is realized when $\Delta=0.7002483$ and 
$t=0.51542$. For both Heisenberg and AKLT chains, the quasiparticle picture well holds
even under the hyperbolic deformation, and the estimated gaps $\Delta$ by 
these finite size scalings are consistent with known values.~\cite{Schollwock:PRB53, 
Okunishi:PRB59, Nakano, Todo:PRL87}

\begin{figure}
\begin{center}
\includegraphics[width=80mm]{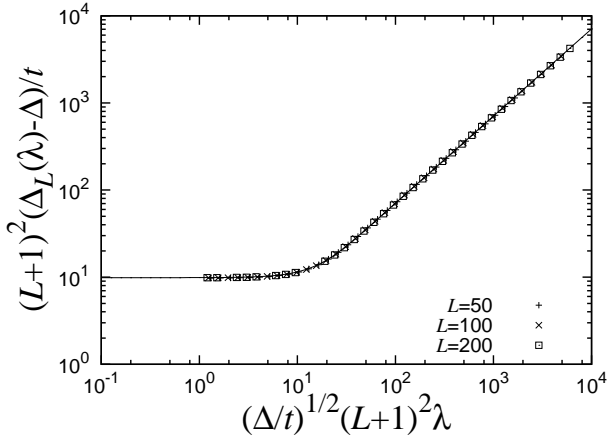}
\end{center}
\caption{Scaling plot for the hyperbolically deformed AKLT chain. The parametrization
$\Delta=0.7002483$ and $t=0.51542$ is used. 
}
\end{figure}
\begin{figure}
\begin{center}
\includegraphics[width=80mm]{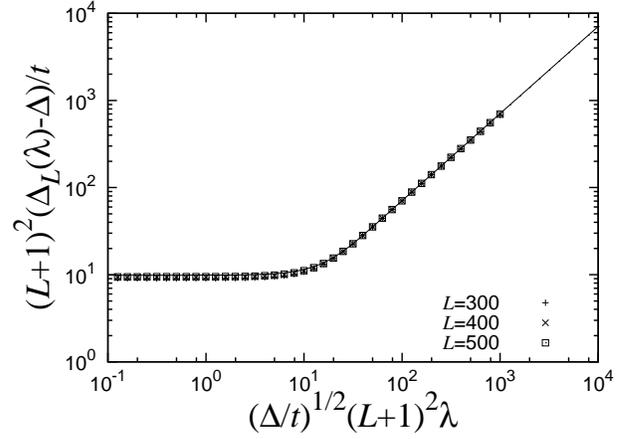}
\end{center}
\caption{Scaling plot for the hyperbolically deformed XXZ chain, where 
$\alpha = 0.8$ and $D = 0$ in Eq.~(4.2). The
parametrization $\Delta=0.20502$ and $t=13.01$ is used. 
}
\end{figure}

Let us check the validity of two-parameter scaling on another model,
the $S = 1$ XXZ spin chain with uniaxial anisotropy. Under the hyperbolic 
deformation, the corresponding Hamiltonian is written as
\begin{eqnarray}
&& \!\!\!\!\!\!\!\!
{\hat H}^{XXZ}_{~}(\lambda)  = 
\sum_{j=-L/2+1}^{L/2-1} \cosh j \lambda  \biggl[
S_j^{X} S_{j+1}^{X} +
S_j^{Y} S_{j+1}^{Y}  \nonumber\\
&& \!\!\!\!\!\!\!\!
+ \alpha \, S_j^{Z} S_{j+1}^{Z} \biggr] 
+ D \!\!\!\! \sum_{j=-L/2+1}^{L/2} \cosh (j-{\textstyle \frac{1}{2}}) \lambda  \,\, 
\left[ S_j^z \right]^2_{~} \, . \nonumber\\
\end{eqnarray}
Figure 3 shows the scaling plot when $\alpha = 0.8$ and $D = 0$. For this case the 
scaling parameters are determined as $\Delta=0.20502$ and $t=13.01$. Figure 4 
is the case when $\alpha = 1$ and $D=0.2$. For this case we obtain $\Delta=0.288240$ 
and $t=10.77$. Again the two-parameter scaling well holds 
for ${\hat H}^{XXZ}_{~}( \lambda )$, and the obtained value of $\Delta$ agrees 
with previous studies.~\cite{Chen:PRB67, Boschi:EPJB35, Ueda:PRB78} 

\begin{figure}
\begin{center}
\includegraphics[width=80mm]{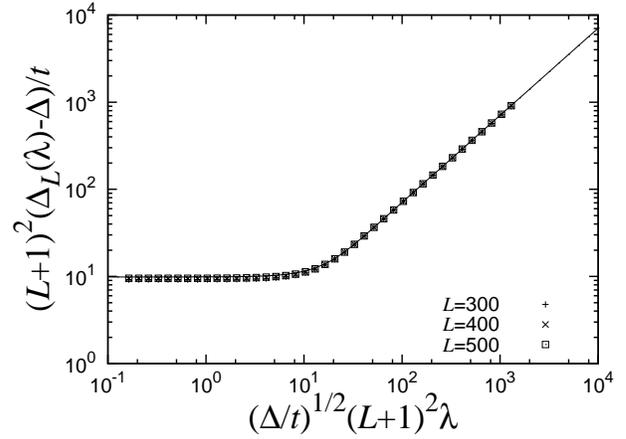}
\end{center}
\caption{Scaling plot for the case $\alpha = 1$ and $D=0.2$ in Eq.~(4.2), under the
parametrization $\Delta=0.288240$ and $t=10.77$.
}
\end{figure}

A profit of the two-parameter scaling is that the effective hopping parameter $t$ is
obtained simultaneously with the gap $\Delta$. It is well known that the spin 
velocity
\begin{equation}
v = \sqrt{2t\Delta} 
\end{equation}
is essential for the thermodynamic property of the spin chains at low 
temperature.~\cite{Sorensen:PRL71}  (See Appendix B.)
From the data in Fig.~1, the velocity at the isotropic Heisenberg point is obtained as 
$v = 2.462$, which is consistent with $v = 2.4691$ reported by S.~Todo.~\cite{Todo:PRL87}. 
At the AKLT point, the estimated value of the velocity from Fig.~2 is $v = 0.84962$. 
The value deviates from $v = 0.825$ reported by K.~Okunishi et al,~\cite{Okunishi:PRB59}
and the clarification of this difference is a remaining problem.

\section{Conclusion}

We have investigated the elementary excitation energy of the hyperbolically deformed $S = 1$ 
spin chains. Two-parameter scaling for the excitation energy $\Delta_L^{~}( \lambda )$
coincides with the effective one-quasiparticle picture in Eq.~(3.2), which is also 
written in the form
of hyperbolic deformation. The fact makes it possible to estimate both the bulk gap 
$\Delta$ and the effective hopping parameter $t$ accurately by introduction of the
hyperbolic deformation.

There are some variations of the non-uniform deformation applied to spin chains. A simple example is
the exponential deformation introduced by Okunishi.~\cite{Okunishi:exp,Okunishi:Kondo} 
Similar to the hyperbolic deformation, the correlation length becomes finite when the 
site-dependent energy scale $f_j^{~} = e^{j\lambda}_{~}$ is introduced. The behavior of the 
excited quasiparticle from the gapped ground state under this {\it exponential deformation} 
is our interest for future study.

\acknowledgements
The authors thank to Koichi Okunishi for valuable discussions.
This work was partly supported by Grants-in-Aid for JSPS Fellows, Grants-in-Aid from the 
Ministry of Education, Culture, Sports, Science and Technology (MEXT) (No.~19540403, 
No.~20340096, and No.~22014012), and the Global COE Program (Core Research and 
Engineering of Advanced Materials-Interdisciplinary Education Center for Materials Science), 
MEXT, Japan.

\appendix
\section{Numerical Details}

We explain several numerical details on the calculation of $E_L^{(0)}( \lambda )$ and 
$E_L^{(1)}( \lambda )$ in Eq.~(4.1) by means of the DMRG method. To simplify the 
formulation without loosing generality, we consider ${\hat H}( \lambda )$ defined by 
Eq.~(2.1). According to the custom in DMRG, let us split ${\hat H}( \lambda )$ 
into three parts
\begin{equation}
{\hat H}( \lambda ) 
= {\hat H}_L^{~}( \lambda ) + {\hat h}_{0,1}^{~} + {\hat H}_R^{~}( \lambda ) \, ,
\end{equation}
where ${\hat H}_L^{~}( \lambda )$ and ${\hat H}_R^{~}( \lambda )$ are, respectively,
the left and the right block Hamiltonians, which are defined as
\begin{eqnarray}
{\hat H}_L^{~}( \lambda ) &=& \sum_{j=-L/2+1}^{-1}  
\cosh j \lambda \,\, {\hat h}_{j,j+1}^{~}
\nonumber\\
{\hat H}_R^{~}( \lambda ) &=& ~~ \sum_{j=1}^{L/2-1}  
~~ \cosh j \lambda ~ {\hat h}_{j,j+1}^{~} \, .
\end{eqnarray}
These block Hamiltonians satisfy a recursion relation~\cite{Ueda}
\begin{eqnarray}
{\hat H}_L^{~}( \lambda ) 
&=& 
\cosh \lambda \,\, {\hat h}_{-1,0}^{~} - {\hat h}_{-2,-1}^{~} \nonumber\\
&+& 
2 \cosh \lambda \,\, H_L^{*} - H_L^{**}( \lambda ) \nonumber\\
{\hat H}_R^{~}( \lambda ) 
&=& 
\cosh \lambda \,\, {\hat h}_{1,2}^{~} - {\hat h}_{2,3}^{~} \nonumber\\
&+& 
2 \cosh \lambda \,\, H_R^{*} - H_R^{**}( \lambda ) \, ,
\end{eqnarray}
where $H_L^{*}$ and $H_R^{*}$ represent the block Hamiltonians for 
the $(L-2)$-site system
\begin{eqnarray}
{\hat H}_L^{*}( \lambda ) &=& \sum_{j=-L/2+1}^{-2}  
\cosh (j-1) \lambda \,\, {\hat h}_{j,j+1}^{~}
\nonumber\\
{\hat H}_R^{*}( \lambda ) &=& ~~ \sum_{j=2}^{L/2-1}  
~~ \cosh (j-1) \lambda ~ {\hat h}_{j,j+1}^{~} \, ,
\end{eqnarray}
and where $H_L^{**}$ and $H_R^{**}$ represent those for $(L-4)$-site systems
\begin{eqnarray}
{\hat H}_L^{**}( \lambda ) &=& \sum_{j=-L/2+1}^{-2}  
\cosh (j-2) \lambda \,\, {\hat h}_{j,j+1}^{~}
\nonumber\\
{\hat H}_R^{**}( \lambda ) &=& ~~ \sum_{j=2}^{L/2-1}  
~~ \cosh (j-2) \lambda ~ {\hat h}_{j,j+1}^{~} \, .
\end{eqnarray}
The recursion relation Eq.~(A.3) is useful when one performs the infinite system 
DMRG iteration for the preparation of the block Hamiltonians to start 
the following finite system DMRG sweeps. It is also possible to perform the
infinite system DMRG method for a uniform spin chain $\lambda = 1$, and 
increase $\lambda$ afterword adiabatically during the following finite system 
DMRG sweeps. 

A special care, which should be taken for the hyperbolically deformed system, is 
the subtraction of the energy expectation value from the Hamiltonian.
Since the factor $\cosh j \lambda$ increases rapidly with respect to $| j |$, the
value of $E_L^{(0)}$ becomes huge when $\frac{L}{2} \lambda$ is large. 
This is a cause of numerical
error. Thus we shift the origin of the energy so that the ground-state energy 
becomes almost zero.  This energy shift can be performed during the
finite size sweeping.  First we obtain the ground state $| \Psi_0^{~} \rangle$ 
diagonalizing the Hamiltonian ${\hat H}( \lambda )$, which is represented as
the super-block Hamiltonian in DMRG formulation. We calculate the 
expectation value $\langle {\hat h}_{j,j+1}^{~} \rangle$ for each bond between 
the {\it active} two sites between the left and the right blocks. Every time we calculate 
$\langle {\hat h}_{j,j+1}^{~} \rangle$, we subtract $\cosh \lambda j \,
\langle {\hat h}_{j,j+1}^{~} \rangle$ from the bond Hamiltonian, and
create the renormalized block Hamiltonians after this subtraction. Within several sweeps 
the numerical error in $\langle {\hat h}_{j,j+1}^{~} \rangle$ almost vanishes, 
and the ground state energy
calculated for the Hamiltonian under the energy shift
\begin{equation}
\hat{H}^{}_{}(\lambda) - \langle \hat{H}^{}_{}(\lambda) \rangle =
\sum_{j=-L/2+1}^{L/2-1} \cosh j \lambda  \,\, \left[ \hat{h}^{}_{j,j+1} 
- \langle \hat{h}^{}_{j,j+1} \rangle \right] 
\end{equation}
becomes zero within the tiny numerical error. When we put $S = \frac{1}{2}$ boundary
spins, we also perform this energy shift for the additional boundary interactions.

\section{Spin Velocity}

We shortly review a physical interpretation of the spin velocity 
$v = \sqrt{2 t \Delta}$ in Eq.~(4.3). 
In special relativity the energy $E$ of a particle that has momentum $p$ and mass $m$ is 
given by $E = \sqrt{ p^2_{~} c^2_{~} + m^2_{~} c^4_{~}}$, where $c$ is the velocity of light. 
Expanding by $p$ one obtains the non relativistic approximation
\begin{equation}
E \sim m c^2_{~} + p^2_{~} / 2m = m c^2_{~} + p^2_{~} c^2_{~} / 2 m c^2_{~} .
\end{equation}
On the other hand, the dispersion of $H^{\rm eff}_{~}$ in Eq.~(3.1) is given by 
$E^{\rm eff}_{~} = \Delta + 2 t ( 1 - \cos k )$, where $k$ is the wave number
of the quasiparticle. Expanding by $k$ one obtains 
\begin{equation}
E^{\rm eff}_{~} \sim \Delta + t k^2_{~} = \Delta + k^2_{~} 2 t \Delta / 2 \Delta.
\end{equation}
It should be noted that both the lattice constant and $k$ are dimensionless. Thus $m c^2_{~}$ 
corresponds to $\Delta$, and $p$ corresponds to $k$. Comparing Eq.~(B$\cdot$1) and Eq.~(B$\cdot$2), 
one finds that the light velocity $c$ corresponds to a dimensionless velocity $v = \sqrt{2 t \Delta}$, which
is nothing but the velocity of the spin wave when the system is gapless.

\end{document}